\documentclass[english,twocolumn,showpacs,preprintnumbers,amsmath,amssymb]{revtex4}
\usepackage[T1]{fontenc}
\usepackage[latin1]{inputenc}
\usepackage{array}
\usepackage{longtable}
\usepackage{graphicx}
\usepackage{amssymb}

\makeatletter

\providecommand{\tabularnewline}{\\}

\usepackage{dcolumn}

\usepackage{bm}

\AtBeginDocument{
  
}

\usepackage{babel}
\makeatother
\begin{document}

\title{Site percolation on planar $\Phi^{3}$ random graphs}

\author{J-P. Kownacki}

\email{kownacki@ptm.u-cergy.fr}

\affiliation{Laboratoire de Physique Th\'{e}orique et Mod\'{e}lisation \\
\
CNRS-Universit\'{e} de Cergy-Pontoise - UMR8089 \\
 2, avenue Adolphe Chauvin, 95302 Cergy-Pontoise Cedex, France}

\date{\today}

\begin{abstract}
\noindent In this paper, site percolation on random $\Phi^{3}$ planar
graphs is studied by Monte-Carlo numerical techniques. The method
consists in randomly removing a fraction $q=1-p$ of vertices from
graphs generated by Monte-Carlo simulations, where $p$ is the occupation
probability. The resulting graphs are made of clusters of occupied
sites. By measuring several properties of their distribution, it is
shown that percolation occurs for an occupation probability above
a percolation threshold $p_{c}$=0.7360(5). Moreover, critical exponents
are compatible with those analytically known for bond percolation.
\end{abstract}


\maketitle

\section{Introduction}

Percolation on regular lattices has been extensively studied, both
analytically and numerically \cite{key-1}. It is now firmly established
that critical exponents, governing scaling laws near percolation threshold,
depend only on the dimension $d$ of the lattice. On the contrary,
percolation thresholds depend on the precise structure of the lattice.
The complete knowledge of site and bond percolation thresholds for
\textit{all} regular lattices with given dimension is a very interesting
challenge, from both theoretical and experimental points of view.
It is also known that there exists an upper critical dimension $d_{c}=6$.
This means that for dimension $d\ge d_{c}$, all regular lattices
belong to the same universality class as a family of regular lattices
containing no loops, called Bethe lattices. This is due to the fact
that the proportion of closed loops in a large regular lattice decreases
with the dimension $d$ and eventually becomes negligeable for large
$d$. 

On the other hand, percolation on \textit{random} graphs is still
an open subject. One class of such random graphs, called complex networks,
containing scale free networks and Erdös-R\'{e}nyi networks, is at
present attracting a lot of interest in physical and mathematical
communities as they are good models for real networks (world wide
web, social networks, $\ldots$) \cite{key-2}. A crucial feature
of complex networks is that closed loops can be neglected for large
graphs. Percolation theory have been recently used to investigate
their intrinsic properties \cite{key-3}.

A radically different family of random graphs have been extensively
studied in the past decades as a non-perturbative regularization of
quantum gravity (see \cite{key-4} for a review). Contrary to complex
networks, they are planar, with closed loops that cannot be neglected.
Moreover, distant vertices are strongly correlated. In some sense,
these graphs are closer to regular lattices than complex networks.
More precisely, they look \textit{locally} like a regular lattice
but are globally very different. This intermediate situation makes
the problem of percolation on these graphs very exciting. Planar $\Phi^{3}$
random graphs, or their dual planar dynamical triangulations, belong
to this family. They are defined in section II of this paper. Their
properties are rather well established now on the ground of a great
amount of analytical and numerical results \cite{key-5}. In particular,
it is known that the Hausdorff dimension of these graphs is $d_{H}=$4
\cite{key-6} with a fractal structure of so-called baby universes
\cite{key-7}. Percolation on planar $\Phi^{3}$ random graphs is
the subject of this paper. In fact, \textit{bond} percolation on planar
$\Phi^{3}$ random graphs has been exactly solved as the limit $q\rightarrow1$
of a $q$-state Potts model, using matrix models \cite{key-8}. In
this article, we study numerically \textit{site} percolation on these
graphs. The main purpose is to measure the value of the percolation
threshold. Moreover, our work is a test of universality between site
and bond percolation for this model.

\section{the model}

\subsection{Random $\Phi^{3}$ planar graphs}

\begin{figure}
\includegraphics[width=7cm,keepaspectratio]{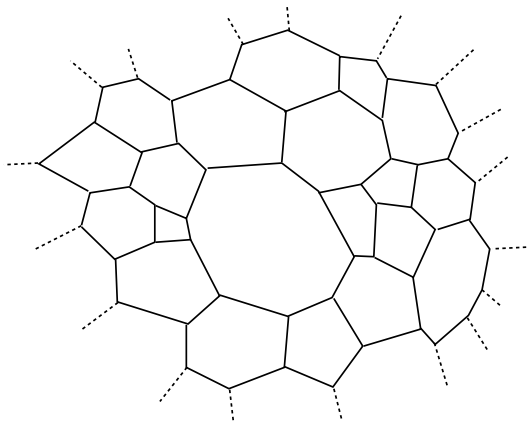}

\caption{A $\Phi^{3}$ graph. \label{phi3_ex}}
\end{figure}

We consider the set of all planar graphs with $N$ trivalent vertices
, \textit{i.e} graphs without boundaries that can be drawn on a sphere
and where each vertex is linked to exactly three neighbors. Morover,
two distinct vertices can be linked by at most one link and no vertex
can be linked to itself (see Fig. \ref{phi3_ex}). %

These graphs are purely topological objects as no length scale is
given here. Such graphs are characterized by their Euler number $\chi=N-N_{l}+N_{f}=2$,
where $N$, $N_{l}$, and $N_{f}$ are respectively the number of
vertices, links and faces. Moreover, local properties of these graphs
imply $2N_{l}=3N$. However, there is no constraint on the size of
a face, \textit{i.e} the number of links surrounding a face, except
that it must be greater than three. Note that, in some cases (degenerate
graphs), some loops contain almost all links of the graph. This set,
called $\Phi^{3}$ planar graphs, is turned into a statistical model
by assigning a Boltzmann weight to each graph. In this paper, each
graph has the same weight. The partition function of this ensemble
of random graphs is written \begin{eqnarray*}
Z_{N} & = & \sum_{G\in\Phi^{3}|_{N}}\,\frac{1}{C(G)}\end{eqnarray*}
 where the sum is over all $\Phi^{3}$ graphs with N vertices as defined
above, and $C(G)$ is a symmetry factor wich avoid the overcounting
of some symmetric graphs ($C(G)$ is almost always equal to one for
large graphs). The size of each face can be seen as a random variable,
with mean value equal to six, whose distribution can be exactly calculated
\cite{key-9}. However, toplogical constraints (planarity) imply strong
correlations between distant faces. As an example, correlation between
ajacent faces follow a modified Aboav's law \cite{key-10}. Another
important feature of these graphs is their Hausdorff dimension, which
has been shown to be $d_{H}=4$ \cite{key-6}. It can be defined as
follows: first define a path between two vertices $v_{1}$ and $v_{2}$
as a succession of adjacent links connecting $v_{1}$ and $v_{2}$
. The length of the path is the number of links in the path and the
geodesic distance between $v_{1}$ and $v_{2}$ is the length of the
shortest path between them. $\left\langle N_{r}\right\rangle _{o},$
the mean number of vertices whose geodesic distance from an arbitrary
vertex $v_{o}$ is lower than $r$, scales (for large $r$) as

\begin{eqnarray*}
\left\langle N_{r}\right\rangle _{o} & \sim & r^{d_{H}}\end{eqnarray*}

\subsection{Site percolation }

\subsubsection{Definition}

\begin{figure}
\includegraphics[width=7cm,keepaspectratio]{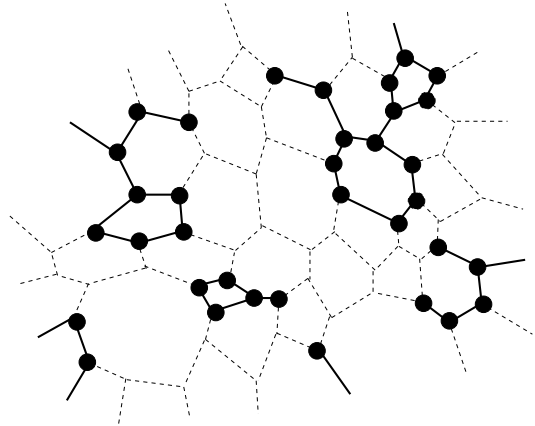}

\caption{Occupied (black bullets) and empty sites on a $\Phi^{3}$ graph.
Dashed links connect empty sites to occupied or empty sites. \label{phi3_ex.2}}
\end{figure}

We now consider the problem of site percolation on these graphs. As
usual, each site (vertex) of a graph \emph{$G$} is randomly occupied
or \emph{}empty, independently of the rest of the sites. \emph{}More
precisely, each site is occupied with probability $p$ or empty with
probability $q=1-p$ ( see Fig. \ref{phi3_ex.2}). Each distribution
of occupied and empty sites on $G$ induces a structure of clusters.
A cluster is a set of occupied sites connected by links of $G$. The
study of average properties of these clusters as $p$ is varied is
the subject of percolation theory. Consider a graph $G\in\Phi^{3}|_{N}$
and a distribution of occupied/empty sites on $G$, denoted $\mathcal{D}$
(for a given occupation probability $p$); suppose that a quantity
$A=A\left[G,\mathcal{D}\right]$ depends on $G$ and $\mathcal{D}$.
The annealed average quantity $\left\langle A\right\rangle (N,p)$
is obtained by first averaging on all distributions $\mathcal{D}$
for a given graph $G$ and, then, by averaging on all graphs $G$.
It can be written:\begin{eqnarray*}
\left\langle A\right\rangle (N,p) & = & \sum_{G\in\Phi^{3}|_{N}}\,\frac{1}{C(G)}\,\sum_{\mathcal{D}}A\left[G,\mathcal{D}\right]\end{eqnarray*}
$\left\langle A\right\rangle (N,p)$ can be seen as the (average)
value of $A$ on a graph picked at random from the $\Phi^{3}|_{N}$
ensemble .

\subsubsection{Cluster distribution}

For fixed $p$, the distribution of the sizes of clusters is described
by the quantity $n(s,p)$, equal to the density of clusters made of
$s$ connected sites. Alternatively, one is interested by all the
moments of the distribution: $K_{n}(p)=\sum_{s}^{'}\, n(s,p)\, s^{n}$,
where the symbol $\sum^{'}$ means that only finite clusters enter
the sums. In practice, only the first moments are usually studied.
More precisely, $K_{o}(p)=\sum_{s}^{'}\, n(s,p)$ is the total number
per site of finite clusters; $K_{1}(p)=\sum_{s}^{'}\, n(s,p)\, s$
is the probability that a site belongs to any finite cluster and $S(p)=\sum_{s}^{'}\, n(s,p)\, s^{2}/\sum_{s}^{'}\, n(s,p)\, s$
is one way to define the mean cluster size.

Other interesting quantities are $s_{1}(p)$, $s_{2}(p)$, $\ldots$,
defined as the sizes of the largest, second largest,$\ldots$ cluster.

\subsubsection{Percolation threshold }

One of the most basic questions is the existence, in the thermodynamic
limit $N\,\rightarrow\infty$, of a percolation threshold $p_{c}$
such that, for $p<p_{c}$, all clusters have a finite size and, for
$p>p_{c}$, there exists at least one spanning cluster. This is formally
expressed by defining a probability $R(p)$ that there exists a spanning
cluster, with $R(p)=0$ for $p<p_{c}$ and $R(p)=1$ for $p>p_{c}$.
The definition of a spanning cluster is not unique but, practically,
it can be seen as a cluster connecting two opposite boundaries on
a lattice. Unfortunately, such a definition does not easily apply
for the family of graphs considered in this paper, as there is no
natural way to define opposite boundaries in these graphs. However,
this can be circumvented by defining $p_{c}$ as the value of $p$
for which the sizes of the second, third, $\ldots$ largest clusters
reach a maximum \cite{key-11}. This property can be understood with
the following heuristic argument: for $p<p_{c}$ , all graphs contain
several finite clusters with comparable size, as soon as $p$ is not
too close to $p_{c}$ . As $p$ is growing, the sizes of the largest,
second largest, third largest $\ldots$clusters are growing until
$p_{c}$ is reached. At this point, most of the clusters merge into
a giant component so that, for $p>p_{c}$, the sizes of the second,
third,$\ldots$ largest clusters dramatically fall down.

\subsubsection{Critical exponents}

There is a profound analogy between percolation and critical phenomena,
where $p_{c}$ plays the role of the critical temperature. The analog
of the order parameter is the density of the spanning cluster, denoted
$\mathcal{P}(p)$ . It represents the probability for a site to belong
to the spanning cluster. It can be defined as $\mathcal{P}(p)=p-\sum_{s}^{'}\, n(s,p)\, s$
since $p$ is the probability for a site to belong to a cluster (finite
or not) and $\sum_{s}^{'}\, n(s,p)\, s$ is the probability for a
site to belong to a finite cluster. For $p<p_{c}$, all clusters are
finite and $\mathcal{P}(p)=0$. For $p>p_{c}$ , $\mathcal{P}(p)$
increases with $p$ and $0<\mathcal{P}(p)\le1$. However, for planar
$\Phi^{3}$ random graphs, the notion of spanning cluster is not appropriate,
and $\mathcal{P}(p)$ is defined as the density of the largest cluster.
At least for $p\sim p_{c}$ and large graphs, both definitions are
expected to coincide.

As in critical phenomena, it is possible to define a correlation length
$\xi(p)$ related to the mean radius of clusters and diverging at
$p=p_{c}$ as $\xi(p)\sim\left(p-p_{c}\right)^{-\nu}.$ Moreover,
most observables are expected to follow scaling laws in the region
$p\sim p_{c}$. We define standard exponents $\alpha,\beta$ and $\gamma$
associated with $K_{o}(p)$, $\mathcal{P}(p)$, $S(p)$ in the critical
region by \begin{eqnarray*}
K_{o}(p) & \sim & \left(p-p_{c}\right)^{2-\alpha}\\
\mathcal{P}(p) & \sim & \left(p-p_{c}\right)^{\beta}\\
S(p) & \sim & \left(p-p_{c}\right)^{-\gamma}\end{eqnarray*}

and, at $p=p_{c}$, $n(s)=n(s,p_{c})$ is expected to scale with an
exponent $\tau$ as\begin{eqnarray*}
n(s) & \sim & s^{-\tau}\end{eqnarray*}

These exponents are expected to follow the scaling relations\begin{eqnarray}
2-\alpha & = & \gamma+2\beta\,=d\nu\nonumber \\
\tau & = & 2+\frac{\beta}{d\nu-\beta}\label{eq:scaling}\end{eqnarray}

Here, $d$ is a parameter characteristic of the model. For regular
\emph{D}-dimensional lattices with $D\le6$, $d=D$ is the standard
dimension of the lattice. For $D\ge6$, i.e. above the upper critical
dimension, $d=6$. For planar $\Phi^{3}$ random graphs, $d=d_{H}=4$,
the Hausdorff dimension defined above. 

The sizes of the largest, second largest, $\ldots$ clusters follow
a scaling law \cite{key-11} \begin{eqnarray*}
s_{1} & \sim & \left(p-p_{c}\right)^{\beta-\nu d}\\
s_{2} & \sim & \left(p-p_{c}\right)^{\beta-\nu d}\\
 & \vdots\end{eqnarray*}

\subsubsection{Theoretical values of critical exponents for bond percolation}

Percolation could be defined by an occupation probability on the links
of the planar $\Phi^{3}$ random graphs. This would give rise to \emph{bond
percolation}. As in critical phenomena, critical exponents are expected
to be universal whereas $p_{c}$ should depend on the details of the
model. So, critical exponents should be the same in bond and site
percolation. Bond percolation on planar $\Phi^{3}$ random graphs
has been exactly solved using a random matrix model formulation \cite{key-8}
, so that exponents $\nu,\alpha,\beta,\gamma$ and $\tau$ are exactly
known in this case \cite{key-12}: \[
\nu=1\,\,,\,\,\alpha=-2\,\,,\,\,\beta=\frac{1}{2}\,\,,\,\,\gamma=3\,\,,\,\,\tau=\frac{15}{7}\]
If universality holds, critical exponents for site percolation should
also be given by these values.

\section{The numerical experiment}

\subsection{The method}

\subsubsection{Nodes removal}

The first scoincidetage is to generate graphs in $\Phi^{3}|_{N}$
. We start from a tetrahedron. Then, one face (triangle) is randomly
chosen, a vertex is added inside this face and linked to the three
vertices of the triangle. This procedure is repeated until a polyhedron
with $N$ faces (triangles) is obtained. This polyhedron is transformed
into a $\Phi^{3}$ graph by duality, \emph{i.e.} each face (triangle)
of the polyhedron is replaced by a vertex linked to three vertices
associated to the three adjacent faces of the triangle. The graph
$G^{(o)}$ thus obtained contains $N$ trivalent vertices and has
the topology of a sphere. Then, starting from $G^{(o)}$, all graphs
in $\Phi^{3}|_{N}$ can be generated by using standard flips of links
called $T_{2}$ moves \cite{key-4}.

Once a graph $G\in\Phi^{3}|_{N}$ has been obtained, one distribution
$\mathcal{D}(G,p)$ of occupied and empty sites is randomly generated.
This is achieved by randomly removing $q\, N$ vertices from $G$,
with $q=1-p$. The $p\, N$ remaining vertices are defined as occupied
vertices on $G$. Then, interesting quantities are measured. Several
distributions are generated this way to obtain average properties
on $G$ for fixed $p$. Then, another graph is generated by a serie
of $T_{2}$ moves and the removal procedure is applied to this new
graph, \emph{etc}. Annealed averages of observables are thus computed.

\subsubsection{Cluster construction and measured quantities}

For each distribution $\mathcal{D}(G,p)$ on a given graph $G$, all
clusters are constructed using a breadth-first search algorithm analog
to Wolff algorithm \cite{key-13} . At step $n-1$ of the algorithm,
suppose $n-1$ clusters $c_{1},c_{2},\ldots,c_{n-1}$ have already
been constructed. All corresponding occupied sites are labeled {}``visited''.
Step $n$ first consists in choosing an occupied site not yet visited.
This site $v_{o}$ is the root of the cluster $c_{n}$ and is now
labeled {}``visited''. It is put in a (empty) list $Q$. The following
procedure is now applied to $Q$ : for each site $v$ in $Q,$ all
occupied and not yet visited neighbors of $v$ on $G$ are added to
$c_{n}$, labeled {}``visited'' and put in $Q$ whereas $v$ is
removed from $Q$. This procedure is repeated until $Q$ becomes empty.
The cluster $c_{n}$ is thus completely constructed. The algorithm
stops when all occupied vertices have been visited.

The size of each cluster - \emph{i.e.} the number of sites in the
cluster - is registred and a histogram of the sizes is built. This
allows to measure the following quantities:

\begin{itemize}
\item $K_{o}(p)$ , the total number of clusters (except the largest one);
\item $n(s,p)$ , the density of clusters of size $s$; 
\item $\mathcal{P}(p)$ , the size of the largest cluster divided by $N$;
for large $N$, this is expected to represent the order parameter
defined above;
\item $s_{2}(p)$ , the size of the second largest cluster ;
\item $K_{2}(p)$ , the second moment of $n(s,p)$; in the critical region,
$K_{2}(p)$ is expected to scale like $S(p)=K_{2}(p)/K_{1}(p)$, the
mean (finite) cluster size .
\end{itemize}

\subsubsection{Finite size scaling}

An important tool when using numerical simulations is the finite size
scaling analysis. It is based on the hypothesis that, for finite systems,
scaling laws are corrected by scaling functions depending on the ratio
between the linear size $L$ of the system and the correlation length
$\xi$ . More precisely, for $L\gg\xi$, the system should not feel
finite size effects, but when the correlation length becomes of order
$L$, finite size should modify scaling laws. For a quantity $O$,
a scaling law $O\sim\left(p-p_{c}\right)^{-z}$ (for infinite size
of the system) can be rewritten $O\sim\xi^{z/\nu}$ as the correlation
length itself scales as $\xi\sim\left(p-p_{c}\right)^{-\nu}$. Then,
when $\xi\sim L$, $O\sim L^{z/\nu}$. This can be summarized by $O\sim L^{z/\nu}\, F\left(\left(p-p_{c}\right)\, L^{1/\nu}\right)$
where $F(x)$ is a (scaling) fonction of the dimensionless ratio $x=\left(p-p_{c}\right)L^{1/\nu}\sim\left(L/\xi\right)^{1/\nu}$
such that $F(x)\rightarrow1$ for $x\sim1$ and $F(x)\rightarrow x^{-z}$
for $x\gg1$. 

Finite size scaling laws enable one to extract the values of critical
exponents by studying the behavior of quantities as the size of the
system is varied. This also defines a finite size percolation threshold:
suppose that the scaling function for $O$ reaches a maximum for $x=x_{o}$.
Then, for fixed $L$, the value of $p$ for which $O$ reaches a maximum
is given by $\left(p-p_{c}\right)\, L^{1/\nu}=x_{o}$. This defines
an effective finite size percolation threshold $p_{c}(L)$ approaching
$p_{c}$ when $L\rightarrow\infty$ as $p_{c}(L)-p_{c}\sim L^{-1/\nu}$
\cite{key-1}. 

For percolation on graphs in $\Phi^{3}|_{N}$ , there is no explicit
linear size as graphs are purely topological. However, the quantity
$N^{1/d_{H}}$, where $d_{H}$ is the Hausdorff dimension, defines
an effective linear size, so that finite size scaling can be written
$O\sim L^{z/\nu}\, F\left(\left(p-p_{c}\right)\, N^{1/\nu d_{H}}\right)$
. The effective finite size percolation threshold $p_{c}(N)$ then
approaches $p_{c}$ as $p_{c}(N)-p_{c}\sim N^{-1/\nu d_{H}}$.

\subsection{Simulations}

We simulated graphs of sizes ranging from $N=400$ to $N=25600$ or
$N=51200$ (according to the measured quantity) vertices. When measuring
$\mathcal{P}(p)$ and $s_{2}(p)$, we considered various values of
$p$ for each size. For each graph $G$ and given $p$ , we generated
$n_{\mathcal{D}}=128$ to $1024$ occupation distributions by the
nodes removal method. $n_{\mathcal{D}}$ was chosen big enough to
minimize its influence on annealed averages. Each simulation consisted
in generating $n_{G}$ graphs in $\Phi^{3}|_{N}$, with $n_{G}$ ranging
from $1024$ to $4096$ for each value of $p$. Each graph was obtained
from the previous one by performing $2000\, N_{l}$ flips ($T_{2}$
moves). We estimated error bars by standard jackknife method. Error
bars are always plotted on the figures below but, most of the time,
they are hidden by the symbols.

\subsection{The results}

\subsubsection{Order parameter}

\begin{figure}
\includegraphics[width=7cm,keepaspectratio,angle=-90]{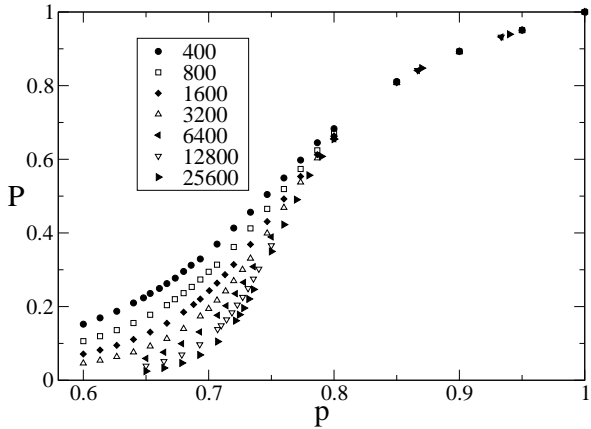}

\caption{Order parameter $\mathcal{P}$ \emph{vs}$p$.\label{P.1}}
\end{figure}

The percolation order parameter $\mathcal{P}$ plotted as a function
of $p$ is shown in Fig. \ref{P.1}, for various values of $N$. We
see that, for low values of occupation probability $p$, $\mathcal{P}$
is close to zero. As expected, $\mathcal{P}=1$ for $p=1$ since all
graphs are connected when no vertex is removed. According to percolation
theory, $\mathcal{P}$ is not zero only in the percolating phase.
As can be seen on the figure, in our model, a percolation transition
is expected to take place for $p\simeq0.7$. However, it is not possible
to get a precise estimate of the threshold value $p_{c}$ with these
data. 

\subsubsection{Percolation threshold $p_{c}$}

\begin{figure}
\includegraphics[width=7cm,keepaspectratio,angle=-90]{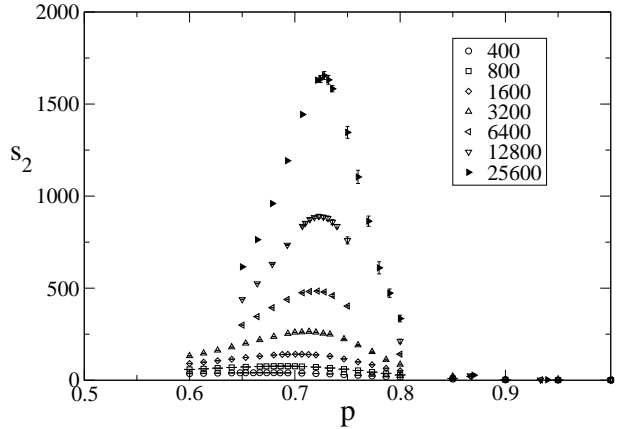}

\caption{ Size of the second largest cluster $s_{2}$ \emph{vs} $p$. \label{S.1}}
\end{figure}

\begin{figure}
\includegraphics[width=7cm,keepaspectratio,angle=-90]{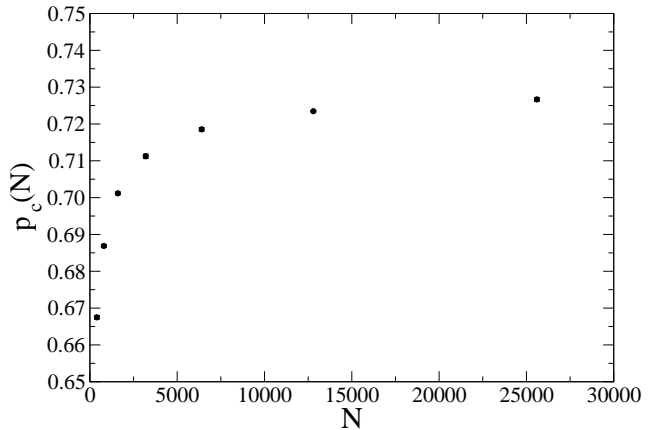}

\caption{Finite size percolation threshold \emph{vs} $N$.\label{pc.1}}
\end{figure}

The percolation threshold $p_{c}$ is determined using the behavior
of $s_{2}$, the size of the second largest cluster. Fig. \ref{S.1}
clearly shows a peak of $s_{2}$ for a value depending on $N$, denoted
$p_{c}(N)$ . This effective percolation threshold $p_{c}(N)$ was
extracted by fitting the peaks of $s_{2}$ with quadratic functions.
The result is plotted in Fig. \ref{pc.1}. As explained above, by
using finite size scaling hypothesis, $p_{c}(N)$ is expected to approach
$p_{c}$ for large $N$ according to \begin{eqnarray}
p_{c}(N) & = & p_{c}+c_{1}N^{-1/\nu d_{H}}\label{eq:threshold_finite}\end{eqnarray}
where $c_{1}$ is a constant. So, we fitted $p_{c}(N)$ with the scaling
law (\ref{eq:threshold_finite}) and we obtained $p_{c}=0.7360(5)$.

\subsubsection{Critical exponents}

\begin{figure}
\includegraphics[width=7cm,keepaspectratio,angle=-90]{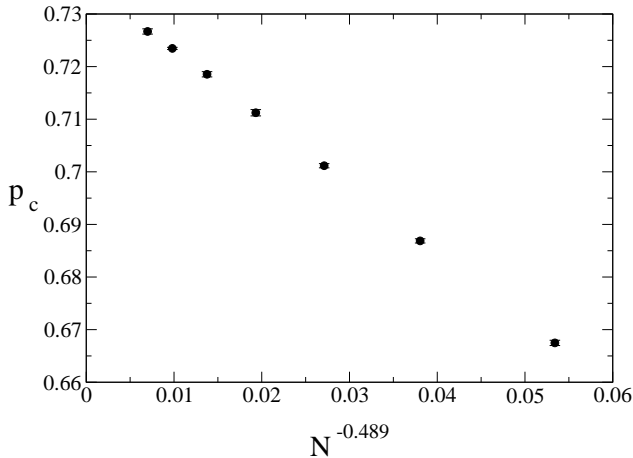}

\caption{ Finite size percolation threshold plotted as a function of $N^{-0.489}$.
\label{fig:pc.2}}
\end{figure}

\begin{itemize}
\item Exponent $\nu$.
\end{itemize}
The fit of $p_{c}(N)$ with the scaling law (\ref{eq:threshold_finite})
allowed us to extract also the value of $1/\nu d_{H}$ . We obtained
$1/\nu d_{H}=0.489(9)$. As can be seen in Fig. \ref{fig:pc.2}, $p_{c}(N)$
is clearly a straight line when plotted as a function of $N^{-0.489}$.

\begin{figure}
\includegraphics[width=7cm,keepaspectratio,angle=-90]{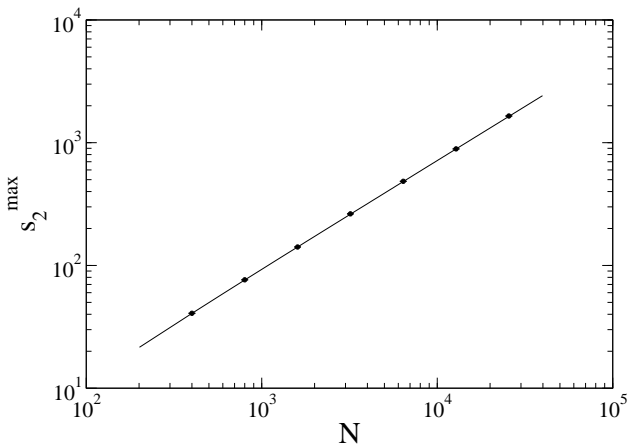}

\caption{Maximum of $s_{2}$ \emph{vs} $N$. The straight line is the best
fit. \label{fig:smax}}
\end{figure}

\begin{itemize}
\item Exponent $\beta$.
\end{itemize}
We also used $s_{2}$ to extract $\beta/\nu d_{H}$: Fig. \ref{S.1}
shows that the peaks of $s_{2}$ are growing with the size $N$. By
fitting the peaks of $s_{2}$ with quadratic functions, we obtained
the value of $s_{2}$ at the maximum, denoted at $s_{2}^{max}(N)$
. Fig. \ref{fig:smax} shows the results. 

By finite size scaling arguments, $s_{2}$ is expected to behave as\begin{eqnarray*}
s_{2} & \sim & N^{1-\beta/\nu d_{H}}\, F\left(\left(p-p_{c}\right)\, N^{1/\nu d_{H}}\right)\end{eqnarray*}
so that $s_{2}^{max}(N)$ scales as\begin{eqnarray}
s_{2}^{max} & \sim & N^{1-\beta/\nu d_{H}}\label{eq:s2max}\end{eqnarray}

By fitting $s_{2}^{max}(N)$ with the scaling form \eqref{eq:s2max},
we obtained $1-\beta/\nu d_{H}=0.880(1)$. The best fit is plotted
in Fig. \ref{fig:smax}.

\begin{figure}
\includegraphics[width=7cm,keepaspectratio,angle=-90]{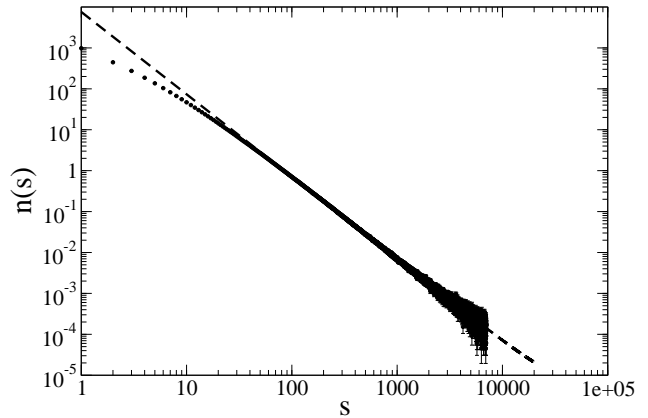}

\caption{Number of clusters of size $s$ (not normalized) for $p=p_{c}$ and
$N=51200.$ The straight line is the best fit.\label{fig:n_s}}
\end{figure}

\begin{itemize}
\item exponent $\tau$ 
\end{itemize}
We measured the number of clusters of size $s$ at $p=p_{c}=0.7360$.
In order to extract the value of $\tau$, we did not use finite size
scaling analysis. Instead, we used the scaling law (for $s\gg1$)
\begin{eqnarray}
n(s) & \sim & s^{-\tau}\label{eq:n_s}\end{eqnarray}
for one value of $N$ large enough to minimize finite size effects.
We chose $N=51200$ and discarded data with $s\gtrsim100$ and $s\lesssim7000$
(because the scaling form is valid for $s\gg1$ and there was not
enough statistic beyond $s=7000$). Then, we fitted the data with
the scaling form \eqref{eq:n_s}. We obtained $\tau=2.0(1)$. The
result is shown in Fig.\ref{fig:n_s}.

\begin{figure}
\includegraphics[width=7cm,keepaspectratio,angle=-90]{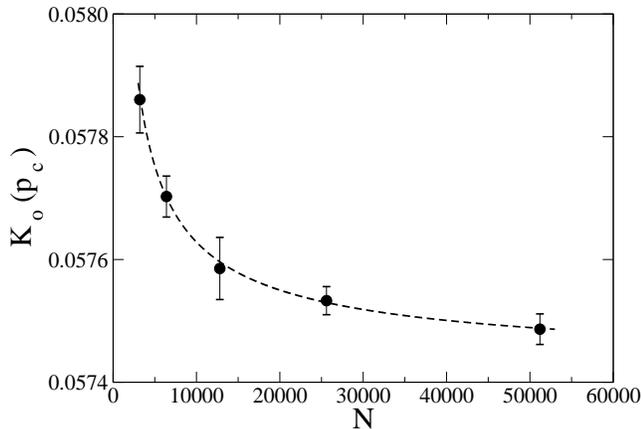}

\caption{Density of total number of cluster measured at $p=p_{c}$. The dashed
curve is the the best fit.\label{fig:NBC}}
\end{figure}

\begin{itemize}
\item Exponent $\alpha$
\end{itemize}
We extracted the value of $\alpha$ by using finite size scaling analysis
applied to the density of total number of clusters $K_{o}(p)$ measured
at $p=p_{c}$ (Fig. \ref{fig:NBC}). The expected scaling law is \begin{eqnarray}
K_{o}(N,p_{c}) & \sim & N^{\left(\alpha-2\right)/\nu d_{H}}\label{eq:nc_tot}\end{eqnarray}

We fitted data with \eqref{eq:nc_tot} and obtained $\left(\alpha-2\right)/\nu d_{H}=-0.7(2)$.
The result is plotted in Fig. \ref{fig:NBC}

\begin{figure}
\includegraphics[width=7cm,keepaspectratio,angle=-90]{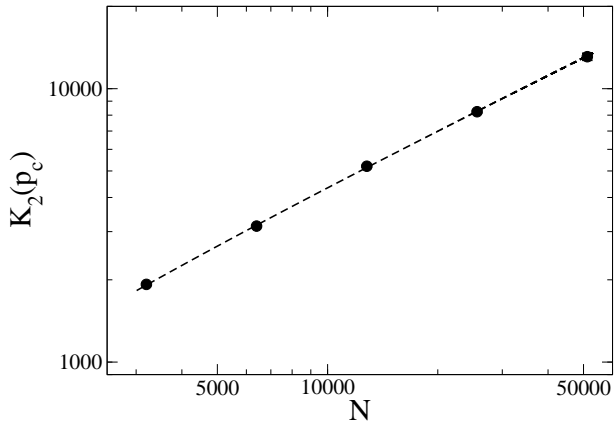}

\caption{Second moment of $n(s,p)$ for $p=p_{c}$. The dashed line is the
best fit.\label{fig:M2}}
\end{figure}

\begin{itemize}
\item Exponent $\gamma$
\end{itemize}
We measured $K_{2}(p)$, the second moment of the distribution $n(s,p)$
for $p=p_{c}$. Fig. \ref{fig:M2} shows the result. As mentioned
above, $K_{2}(p)$ is expected to scale as $S(p)$, the mean size
of finite clusters, at least in the critical region. Finite size scaling
law for $K_{2}(p)$ is then\begin{eqnarray}
K_{2}(N,p_{c}) & \sim & N^{\gamma/\nu d_{H}}\label{eq:M2}\end{eqnarray}

We extracted $\gamma$ by fitting data with \eqref{eq:M2} and we
obtained $\gamma/\nu d_{H}=0.67(3)$

\subsubsection{Remark on the results}

The determination of the percolation threshold using the maximum of
the second largest cluster gives rather accurate value for $p_{c}$.
This is partly due to the fact that the parameter $p_{c}$ in the
scaling law \eqref{eq:threshold_finite} is independent of the precise
approach of $p_{c}(N)$ to its infinite size limit. 

The situation is rather different for the determination of critical
exponents. In Table \ref{tab:crit_expo}, measured and theoretical
values of critical exponents considered in this paper are summarized.
In addition, combinations entering scaling relations \eqref{eq:scaling}
are given in Table \ref{tab:Test-of-scaling}. As can be seen, the
value of $1/\nu d_{H}$ obtained by simulations is not compatible
with the theoretical value $1/4$. In fact, it is well known that
the Hausdorff dimension of planar $\Phi^{3}$ random graphs is very
sensitive to finite size effects and can be extracted only for large
lattices using sophisticated scaling variables in finite size scaling
analysis \cite{key-14}. However, this does not mean that simulations
are unable to take account of the fractal structure of these graphs.
The remaining critical exponents are compatible ($\beta/\nu d_{H}$,$\tau$)
or marginally compatible ( $\left(\alpha-2\right)/\nu d_{H}$ , $\gamma/\nu d_{H}$)
with the theoretical values. However, error bars take into account
neither the error on $1/\nu d_{H}$ nor the uncertainty in determining
$p_{c}$. Moreover, logarithmic corrections to scaling should be considered
to obtain accurate values of the exponents. 

\begin{table}
\begin{longtable}{|c|c|c|c|c|c|}
\hline 
Exponent&
$1/\nu d_{H}$&
$\left(\alpha-2\right)/\nu d_{H}$&
$\beta/\nu d_{H}$&
$\gamma/\nu d_{H}$&
\multicolumn{1}{c|}{$\tau$}\tabularnewline
\hline
\hline 
Simulation&
$0.489(9)$&
$-0.7(2)$&
$0.120(1)$&
$0.67(3)$&
$2.0(1)$\tabularnewline
\hline 
Theory&
$0.25$&
$-1$&
$0.125$&
$0.75$&
$15/7$\tabularnewline
\hline
\end{longtable}

\caption{Theoretical and measured values of critical exponents. \label{tab:crit_expo}}
\end{table}

\begin{table}
\begin{tabular}{|c|c|c|}
\hline 
Combination&
$\left(\gamma+2\beta\right)/\nu d_{H}$&
$2+\frac{\beta/\nu d_{H}}{1-\beta/\nu d_{H}}$\tabularnewline
\hline
\hline 
Simulation&
$0.91(3)$&
$2.136(1)$\tabularnewline
\hline 
Theory&
$1$&
$\frac{15}{7}$\tabularnewline
\hline
\end{tabular}

\caption{Test of scaling relations between critical exponents\label{tab:Test-of-scaling}}
\end{table}

\section{Conclusion}

The first important fact is to notice that critical exponents and
scaling relations obtained by simulations are globally compatible
with the expected theoretical values calculated for bond percolation.
This gives confidence in the extraction of the (unknown) site percolation
threshold. This also confirms universality between site and bond percolation
for this model.

However, the main result of this paper is the value $p_{c}=0.7360(5)$
for site percolation on planar $\Phi^{3}$ random graphs. It is greater
but not very far from the threshold value for site percolation on
the honeycomb lattice $p_{c}(honeycomb)=0.6962...$, which is the
simplest regular trivalent lattice. On one hand, this means that,
to some extent, planar $\Phi^{3}$ random graphs and honeycomb lattices
look alike: more precisely , they \emph{locally} look alike. In contrast,
trivalent Bethe lattices are neither locally equivalent to honeycomb
nor to planar $\Phi^{3}$ random graphs, so that their percolation
thresholds are very different, $p_{c}(3-\mbox{Bethe lattice })=1/2$.
On the other hand, as $p_{c}$ is greater for planar $\Phi^{3}$ random
graphs than honeycomb lattices, percolation is easier on a pure hexagonal
lattice than on the planar $\Phi^{3}$ random graphs. The reason is
that on these latter graphs, there are regions called baby universes
(B.U.) connected to the rest of the graph by very small boundaries
called necks \cite{key-7}. Moreover, B.U. can grow on other B.U.,
giving a fractal (self-similar) structure to the graph. So, for a
given occupation probability $p$, the probability that a given B.U.
belongs to a giant connected cluster is proportional to the probability
that at least one vertex of its boundaries is occupied. This is small
compared with the probability that, on a pure honeycomb lattice, a
given region is a part of a giant cluster. This fractal structure
of B.U. is also the main feature that makes honeycomb lattice and
planar $\Phi^{3}$ random graphs globally different at long distance,
so that their critical exponent are different. 

It would be interesting to study in more details the connections between
baby universes and percolation transition. In particular, a non-uniform
occupation probability, depending for instance on the local curvature
or on the structure of B.U., could shed light on this problem. The
role of B.U. could also be studied by real-space renormalization group
analysis. As mentioned above, planar $\Phi^{3}$ random graphs have
a hierarchical structure that makes them look like trees of baby universes.
It is possible to use this self-similarity of planar $\Phi^{3}$ random
graphs with respect to B.U. to perform a real-space renormalization
group transformation \cite{key-15} by replacing each baby universe
of last generation \emph{(i.e.} a B.U. with no further B.U. growing
on it) by one supersite \cite{key-1}. Then, if the corresponding
last generation B.U. (including its boundary) contained a spanning
cluster, the supersite is occupied. This defines an occupation probability
$p'$ for the supersite as a (complicated) function of $p$, the occupation
probability of the original graph. 

It should also be noticed that the value of $p_{c}$ found here is
comparable with high values found on Archimedean lattices \cite{key-16}.
It would be interesting to understand if planar $\Phi^{3}$ random
graphs share common \emph{local} characteristics with Archimedean
lattices.

\end{document}